# A novel strategy for multi-resource load balancing in agent-based systems

## Leszek Śliwko* and Aleksander Zgrzywa

Institute of Applied Informatics,
Wroclaw University of Technology, Wrocław, Poland
E-mail: leszek.sliwko@pwr.wroc.pl
E-mail: aleksander.zgrzywa@pwr.wroc.pl
*Corresponding author

**Abstract:** The paper presents a multi-resource load balancing strategy which can be utilised within an agent-based system. This approach can assist system designers in their attempts to optimise the structure for complex enterprise architectures. In this system, the social behaviour of the agent and its adaptation abilities are applied to determine an optimal setup for a given configuration. All the methods have been developed to allow the agent's self-assessment. The proposed agent system has been implemented and the experiment results are presented here.

**Reference** to this paper should be made as follows: Śliwko, L. and Zgrzywa, A. (xxxx) 'A novel strategy for multi-resource load balancing in agent-based systems', *Int. J. of Intelligent Information and Database Systems*, Vol. X, No. Y, pp.000–000.

**Biographical notes:** L. Śliwko received his MSc in Computer Science from Technical University of Wrocław in 2005. He is currently employed in Volvo IT as an Application Designer. He is also a Sun Certified Java Programmer and member of BEA Community. His professional experience includes designing of J2EE applications, evaluating the existing system architectures and financial data migrations. His current research (PhD studies) focuses on information systems and intelligent technologies for information retrieval and tasks balancing and includes aspects of agent-based computing. He is a Guest Reviewer of *Int. J. Intelligent Information and Database Systems*.

Aleksander Zgrzywa received his MSc (1970) and PhD (1976) in Computer Science from Wroclaw University of Technology, Poland. He is Professor of Computer Science and Head of the Information Systems Department. His research includes efficiency of the network information systems, application of the AI, e.g., knowledge acquisition in an agent system, negotiating multi-agent systems using Petri nets, Bayesian network approach to information retrieval, applying artificial neuron network for construction of user profile, and using the queuing network to evaluate of the network information systems' efficiency. He is the member of Polish Information Processing Society, IEEE Computer Society and Wroclaw Scientific Society.

# 1 Introduction

Resource management has been an active research area for a considerable period of time. By deploying mobile agent environments, we can achieve some good global system performance (Nguyen and Sliwko, 2005; Nguyen et al., 2006), dynamic adaptation capability (Kim et al., 2004) as well as robustness and fault-tolerance (Denning, 1976; Randell et al., 1978; Xu and Wimsaldonado, 2000). We can also obtain several other benefits such as more cost-effective resource planning (Buyya, 1999), a reduction of network traffic (Kim et al., 2004; Montresor et al., 2003), autonomous activities of the agents (Goodwin, 1995) and decentralised network management (Yang et al., 2005).

Agents can also assist system designers in their attempts to optimise the structure for even complex enterprise architectures. In presented strategy, the social behaviour of the agent and its adaptation abilities (Franklin and Graesser, 1996; Goodwin, 1995) are applied to determine an optimal setup for a given system configuration. Moreover, the algorithm can also be applied to dynamically manage the existing system which utilises several different resources; it can also increase the capacity of any of the networked computers as it assures that each device has a remote access to all resources available in the network (Yang et al., 2005).

This paper is a straight continuation of 'Multi-resource load optimisation strategy in agent-based systems' paper (Sliwko and Zgrzywa, 2007), presented on the *1st KES Symposium on Multi-Agent Systems Conference* in Wrocław. The previously demonstrated algorithm, which is referred to as the KESAMSTA07 strategy, has been extended and also the Java multi-agent system balancer (JMASB) framework went under heavy development. Therefore, we strongly advise the reader to get familiar with our previous work (Nguyen and Sliwko, 2005; Sliwko and Nguyen, 2007; Sliwko and Zgrzywa, 2007) as this paper often refers to it. However, the system basics and the methodology are presented as well in this article.

All the methods have been developed to allow the agent's self-assessment. The proposed schema has been implemented and the experiment results are presented here; additionally, a few other approaches are underlined.

The paper is organised as follows: in Section 2, we describe the idea of utilising agent-based systems in industry standard enterprise architectures. Section 3 provides our standpoint on the multi-resource management through load balancing with working principles explained in details. Section 4 is the most interesting as it presents the results of our experiment compared to various, well-known algorithms. Finally, Section 5 concludes our work and summarises our research.

# 2 Utilising agent-based systems in enterprise architectures

Nowadays, within the globalisation process, the computer system had to become world-wide operating and more powerful (Kesselman and Kesselman, 1999). The amount of data processed in world-wide systems has exceeded by far everything that could be thought of only two decades ago when the first computer system architectures were designed. To process such a large data collection, computer systems need to become more scalable (He, 2000) and portable (Craswell et al., 1997; Wurz and Schuldt, 2005).

This has involved a considerable number of problems but also initiated the process of designing more flexible architecture schemas where the application does not consist in a single atomic entity, but it contains many other components (Alur et al., 2003; Heineman and Councill, 2001; Qu et al., 2004) which could be spread among many environments, being connected only by communication channels.

Among the most well known and developed architectures are Suns Java Enterprise Edition (Alur et al., 2003), Microsoft.NET Framework (Thai and Lam, 2002) and Oracle SOA Suite (Chu et al., 2006). These architectures introduce the concept of components. A component is a module, usually a part of a code or a configuration which can exist on its own (Heineman and Councill, 2001; Szyperski et al., 2002). Such an entity performs programmed actions while hiding real data flow schemas inside. This involves providing an interface which acts as an abstract barrier that the entity creates to yield its functionality to the outside without giving out any details of its implementation. This schema stores away designed decisions that are most likely to change (Parnas, 1972). Thus, other parts of the program are protected in case the component is redesigned.

Such an approach, in turn, requires more effort on the part of the creator of applications, but at the same time, it enables the system to be spread among many networked machines as long as we secure a communication channel for all the system components (Alur et al., 2003; Chu et al., 2006; Thai and Lam, 2002). The enterprise architecture enables the system to utilise a resource of many networked machines. This not only introduces such features as fault tolerance (Randell et al., 1978), but also allows computer systems to be more cost efficient (Alur et al., 2003; Kesselman and Kesselman, 1999; Thai and Lam, 2002) by securing a remote access to all resources available in the network (Yang et al., 2005) for each networked machine.

The next step seemed logical – if we have more than one machine, why not introduce some load management? This has initiated yet another research division. Several systems were designed (Ho and Leong, 2000; Johansson et al., 2002; Montresor et al., 2003), beginning with random task allocation to very complex and specialised industry robots (Johansson et al., 2003).

Nowadays, computer networks often contain applications which can simultaneously use resources of a number of computer environments in order to spread tasks to several machines (Yang et al., 2005). Some applications use the central server (Nguyen and Sliwko, 2005) while others can manage the load asynchronously (Kim et al., 2004; Montresor et al., 2003). The first schema is usually easier to maintain, but the second one often proves more reliable.

The introduction of agent technologies (Franklin and Graesser, 1996) has enabled designers to use a very specific feature which is code migration (Cabri et al., 2006; Craswell et al., 1997). It basically means that not only the data can be moved to another sever, but also the code processing it. Agent frameworks usually support one kind of code migration: weak or strong mobility (Cabri et al., 2006). The weak code migration does not allow any migration of the execution state of methods, i.e., it allows an agent to carry its code and data, but the agent does not carry its state (Wang et al., 2001). The Aglets framework (Cabri et al., 2006; Craswell et al., 1997) provides event-based code callbacks, i.e., the method will be executed after the object initialisation, before the object dispatch, etc. This violates a number of software design principles and then the system developer has to use quite an unnatural programming style (Wang et al., 2001).

Agent-based systems often rely on decentralised architecture (Jones and Crickell, 1997; Kesselman and Kesselman, 1999; Shi et al., 2006) as being more reliable. However, resources are handled through much more complex algorithms (Yang et al., 2005); negotiation schemas are often required for distributed architecture to hold a good level of performance (Johansson et al., 2003).

The agents' migration feature inspired the idea of load balancing through the migration of users' tasks to various networked machines which have spare CPU cycles. This, in turn, enables the users not only to utilise fault tolerance schemas (Xu and Wimsaldonado, 2000) but also to have an inexpensive access to high-end computational capacities on the internet (Kesselman and Kesselman, 1999).

However, the main drawback of the agent migration is the fact that it is also a process which consumes time, resources (Dillenseger and Hazard, 2003) and renders the system inoperable during the task redeployment (Cabri et al., 2006; Wang et al., 2001).

Sometimes, redeploying a working task can be a very complex process. To illustrate this, let us take a job which synchronises several databases. Jobs like this are done through transactions (Alur et al., 2003; Thai and Lam, 2002). We usually begin a few transactions and complete them only after all databases have reported the finish status. Any migration of the task which executes actions like aforementioned can be not only extremely hard to perform, but can also have a big impact on the performance of the whole system. However extreme it seems to be, such a situation is very likely to occur, especially in case of a company which runs several systems sharing some of the data (Chu et al., 2006).

Offering resource management and load balancing features in distributed architectures is a way to enhance the service reliability and hold a good global performance through these various strategies. Job scheduling has been an active research area for a considerable period of time. There exist a number of projects and proposed schemas (Jones and Crickell, 1997), i.e., Messor (Montresor et al., 2003), Traveller (Xu and Wimsaldonado, 2000), LBMA (Yang et al., 2005) and SCARCE (Ho and Leong, 2000). However, most of them focus on achieving optimal utilisation of exclusively one resource and do not regard the costs of the migration process.

In our previous article (Sliwko and Zgrzywa, 2007), we employed the same idea of utilising all the available resources in an optimal way through resource collection comparisons. We designed our strategy as not deterministic as it often tends to stick to the second optimal solution. We demonstrated that this approach can be applied in planning a strategy for the resource utilisation in distributed systems, both as an online- and offline-strategy for solving load balance problems.

In this paper we would like to introduce an enhanced version of algorithm, which behaves better than the strategy presented in Sliwko and Zgrzywa (2007). We also introduced a simple pattern of tasks migration costs, which acts as the objective function (Moura, 2002; Schirmer, 1995) which quantifies the optimality (Cormen et al., 1990; Goldberg, 1989) of the proposed solution.

## 3 The resource management through load balancing

We have previously (Nguyen and Sliwko, 2005; Sliwko and Zgrzywa, 2007) demonstrated that the agent-based resources balancing algorithm is able to hold a good load balance state with various kinds of resources utilised. The experiment (Sliwko and

Zgrzywa, 2007) shown that the strategy could be also applied to unload an overloaded system node within a few cycles needed to spread tasks to other nodes.

The presented schema consists of nodes and agents. Every agent serves one purpose which is called a job. Agents' jobs require resources which are provided by the node. Every node has a certain amount of different resources available. In this paper, it is called the 'available resources' set. To simplify the definition, both the 'resources needed' by the agents' job and the 'resources available' on the node are described by the sets of these resources. There exist several types of resources which can be utilised by the agent. The number of defined resources is potentially unlimited but in our experiment we use only their two kinds. We name them as CPU and memory, which means respectively that the CPU usage agent will cause and the memory agent will allocate to perform its task.

Both introduced strategies share the same idea of optimal resource usage, but the schema presented in this paper introduces also a simple pattern of the 'task migration cost'. Every task has its integer cost value assigned which is an abstract representation of the impact the 'task migration' will have; e.g., this can be understood as the number of hours a developer has to spend to reconfigure a certain service component in another computer environment. We consider this value constant, i.e., a migration of a certain task to any node will cause the same impact throughout the whole global system. Nowadays enterprise applications are deployed in prepared environments, called 'containers' (Alur et al., 2003; Thai and Lam, 2002). The deployment process shall be standardised and automated, regardless of vendor – in most cases it is just enough to drop a file in certain directory and the application server will take care of the proper allocation itself. Thus we assume the amount of work required to initiate the same task on different environments is the same or vary very little. Also the existing method of accessing components is fully transparent and usually done via 'remote interfaces' or 'proxy' object pattern (Alur et al., 2003), thus there is no migration cost dependency; e.g., in case of J2EE architecture, the components references are 'catalogued' by JNDI service (Alur et al., 2003). Another popular approach is to utilise 'service locator' (Alur et al., 2003) pattern, which adds a cashing capability and hides the creation details of a component.

The main task of the system now is not only to find a stable system configuration where no resources on nodes are overloaded, but also to minimise the sum of all required jobs migrations, which we call the 'system transformation cost'.

### 3.1 Problem formulation

Lets define $\Lambda = (\tau, \eta, \psi, a, r, c)$ as our problem space and our system as a twice $(\Lambda, \mu)$. In the 'd-resource system optimisation problem', we receive a set $\tau$ of $l$ mobile tasks: $\tau = \{t_1, t_2, ..., t_l\}$ and a set $\eta$ of $m$ fixed nodes: $\eta = \{n_1, n_2, ..., n_m\}$. We call $\mu : \tau \to \eta$ as a 'task assignment' function, i.e.: every task has to be assigned to the node.

We also consider:

- $\psi = \{i_1, i_2, ..., i_d\}$ as a set of all different kinds of resources; e.g., for $d = 3$ we could define $\psi = \{CPU, memory, network\}$

- $a : \psi \times \eta \to N \cup \{0\}$ as a fixed 'available resources' on the nodes; i.e., $a_i(n)$ is the available level (integer value) of a resource $i$ on the node $n$

- $r : \psi \times \tau \rightarrow N \cup \{0\}$ as a fixed 'required resources' for tasks; i.e., $r_i(t)$ is the required level (integer value) of a resource $i$ of task $t$
- $c : \tau \rightarrow N \cup \{0\}$ $c(t)$ can mean the amount of hours developer has to spend deploying task $t$ on the node.

For every node $n \in \eta$ we define a set $A_n = \{t \in \tau : \mu(t) = n\}$ of all tasks assigned to the node $n$. We also define $f : \psi \times \eta \rightarrow N \cup \{0\}$ as a 'remaining resources' on the nodes:

$$f_i(n) = a_i(n) - \sum_{t \in A_n} r_i(t).$$

We consider system $(\Lambda, \mu)$ as 'stable', iff:

$$f_i(n) \geq 0, \text{ i.e.: } \sum_{t \in A_n} r_i(t) \leq a_i(n), \text{ for every } n \in \eta, i \in \psi \quad (1)$$

Otherwise system $(\Lambda, \mu)$ is 'overloaded'.

Each task $t$ is initially assigned by 'task assignment' function $\mu_0$ to some node $n$; during the 'system transformation' $(\mu_0 \rightarrow \mu_1)$ task $t \in \tau$ can be reassigned to any different node $n \in \eta$. Process of moving the task to different node is called here 'task migration' and this feature generates 'task reassigning cost':

$$c_{(\mu_0 \rightarrow \mu_1)}(t) = \begin{cases} 0, & \mu_0(t) = \mu_1(t) \\ c(t), & \mu_0(t) \neq \mu_1(t) \end{cases}$$

$(\mu_0 \rightarrow \mu_1)$ has its 'system transformation' cost:

$$c_{(\mu_0 \rightarrow \mu_1)} = \sum_{t \in \tau} c_{(\mu_0 \rightarrow \mu_1)}(t) \quad (2)$$

Consider initial 'task assignment' $\mu_0$; 'task assignment' $\mu^*$ is optimal for $\mu_0$, iff $\mu^*$ renders system $(\Lambda, \mu_0)$ 'stable' and:

$$c_{(\mu_0 \rightarrow \mu^*)} \leq c_{(\mu_0 \rightarrow \mu)}, \text{ for every 'stable' system } (\Lambda, \mu).$$

N.b.: when $(\Lambda, \mu_0)$ is 'stable' for initial 'task assignment' $\mu_0$, the 'system transformation cost' equals 0 as it is considered optimal.

### 3.2 Working principles

In computer science, an optimisation problem is a problem of finding the best from all feasible solutions (Cormen et al., 1990; Morton and Mareels, 2000). In the complexity theory, problems are usually phrased as decision problems (Moura, 2002; Schirmer, 1995). For each optimisation problem, there is a corresponding decision problem that asks for yes or no answer to whether there is a feasible solution to some particular measure (Schirmer, 1995).

Nondeterministic polynomial-time hard (NP-hard), represents a class of problems which are 'at least as difficult as problems in NP' (Schirmer, 1995). NP-hard problems may be of any type: search problems, decision problems, optimisation problems or feasibility problems (Schirmer, 1995). Discrete optimisation problems are generally NP-hard problems (Cormen et al., 1990; Garey and Johnson, 1979; Ibaraki and Yagiura, 2004). A defined problem as stated above also belongs to the NP-hard problem class. The proof is out of the scope of this paper; however, our definition can be reduced from a generalisation of the d-dimensional BIN PACKING problem (Epstein and van Stee, 2005; Yoshiharu et al., 2004).

NP-complete problems can be solved by means of exhaustive search (Ueda, 1986). It is now commonly believed that P ≠ NP (Cormen et al., 1990; Moura, 2002; Schirmer, 1995; Yagiura and Ibaraki, 2001). It is rather unlikely that there can ever be any efficient (polynomial time) exact algorithms able to solve NP-hard problems. When the size of that problem instances grows, the number of iterations needed to solve the problem becomes tremendous (Schirmer, 1995); in this case the best we can hope for are super-polynomial time algorithms (Woeginger, 2003).

A fundamental goal in computer science is to provide an algorithm which determines an optimal solution in acceptable time. A heuristic algorithm usually gives up the first feature in order to finish within the satisfactory timeframe (Cormen et al., 1990; Yagiura and Ibaraki, 2001). Generally speaking, it is able to find quite a good solution, but there is no proof that the result could not be better or that the solution found by the heuristic algorithm will be feasible at all. However, heuristic algorithms are nowadays considered one of the most practical approaches to combinatorial optimisation problems (Ibaraki and Yagiura, 2004; Yagiura and Ibaraki, 2001).

### 3.3 Solution details

In this research, our intention is to design an effective strategy that finds a close-to-optimal solution within the acceptable timeframe. As we will show later, our initial results are rather promising.

The main drawback of the previously demonstrated KESAMSTA07 strategy (Sliwko and Zgrzywa, 2007) was the effect we called 'flickering'. Certain setups often resulted in agents rapidly migrating to different nodes, sometimes even within the same cycle. This 'sometimes' caused the system to waste precious iterations on purposeless migrations. Our previous simulation did not involve any 'task migration cost', however, such an action could cause a heavy resource usage in a real system (Cabri et al., 2006; Wang et al., 2001).

Our later research targeted this problem, which resulted in several solutions. The social agent behaviour schema proved to be the best in class; the agent properties such as autonomy, adaptation and communication abilities (Franklin and Graesser, 1996; Goodwin, 1995; Xu and Wimsaldonado, 2000) could be exploited in a very useful manner in many systems.

In our case, each agent represents certain task $t \in \tau$ and therefore fixed 'required resources' $r(t)$ are needed for an agent. Agent is assigned to the node and each node $n \in \eta$ provide only a limited level of 'available resources' $a(n)$. To satisfy (1), nodes in the system must have no more than 'available resources' $a(n)$ allocated. Agent can be reassigned to different nodes $n \in \eta$, that adds a value of 'task migration cost' $c(t)$ to 'system transformation cost' $c_{(\mu_0 \to \mu_1)}$. The main goal of our strategy is to minimise the 'system transformation cost' $c_{(\mu_0 \to \mu_1)}$ to satisfy (2).

The agents cooperate in small groups on nodes to select the best candidate agent for migration. The agent cooperation schema is based on certain evaluation points the agents accredit themselves; we did not utilise any more sophisticated selection strategies such as voting or negotiation schemas (Ho and Leong, 2000) for instance. More complex strategies shall be considered in our future work and we also reckon an option to introduce experience gathering schemas to agents' AI rules.

The IJIIDS08 strategy can be visualised as follows in the UML sequence diagram:

**Figure 1** Sequence diagram of the IJIIDS08 strategy

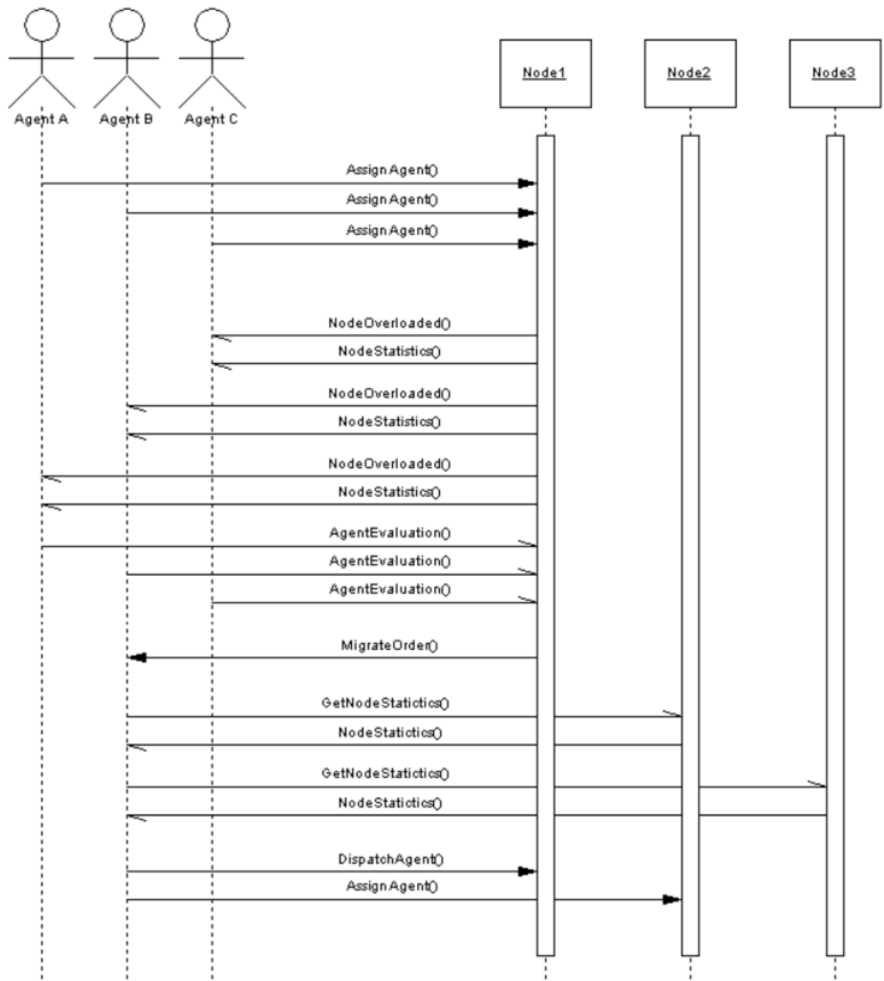

Initially, agents A, B and C assign [*AssignAgent()* message] themselves to Node 1. Such a setup results in Node 1 being overloaded; the node detects this and sends a message to the assigned agents to inform them about the overload [*NodeOverloaded()* message]. Immediately, the node follows with current node statistics [*NodeStatistics()* message] such as resource usage, resource availability, etc. Note that both messages are sent asynchronously.

The agent is a key technology during the next stage. Each agent determines its usefulness on a current node. It considers resources' usage levels, focusing on overloaded resources first. This helps the agent to estimate the real impact its tasks have on the resources of the current node. We have also implemented basic AI elements into the agent. The agent basically 'wants' and 'doesn't want' to do certain actions to a certain degree. Generally speaking, an agent 'doesn't want' to leave its mother node. The level of its resoluteness depends on the 'task migration cost' the agent is carrying with him. The higher 'task migration cost' the less likely the agent is to move to another node. Allowing the agents to select the candidate agent for migration is a major enhancement over the previously presented KESAMSTA07 strategy (Sliwko and Zgrzywa, 2007), where agents were selected based only on their 'required resources' levels. Assigning the percentage chance for an agent migration, instead of selecting the lowest 'migration cost' agent, helped us to avoid the 'flickering' effect (when agents were migrated to different nodes rapidly sometimes even within the same cycle) almost completely.

Besides, an agent that has been previously migrated 'wants' to be migrated again. At this stage, the destination node is still to be decided on, but the agent increases its chances to return to the start node eventually.

Each agent (in this case, agents A, B and C) sends a message about its willingness to migrate to the node. It is represented by an integer value [message *AgentEvaluation()*].

Then the node selects a candidate agent to be migrated during the next stage. The selection is based on the agent opinion:

**Algorithm**  Find a candidate agent for migration

---

Input:

  All agents on this node $A = \{a_1, a_2, ..., a_k\}$

  Agent evaluations (as integer values) $O = \{o(a_1), o(a_2), ..., o(a_k)\}$

  @*resultSignificance* = 1.02

  (*this parameter is adjustable, higher values can result in never selected agents, too low values can result in agents selection being too random*)

Output:

  Agent $a^*$, which will be migrated in the next stage

Begin

  1  For each agent *a* in *A*:

    1.1  If $o(a) \notin O$ then

      1.1.1  Initialise $o(a) = 0$

      1.1.2  Add *o(a)* to *O*

      (*in case of agent a did not return his opinion in time, the o(a) equals 0*)

---

**Algorithm**  Find a candidate agent for migration (continued)

| | | |
|---|---|---|
| | 1.2 | Process $o(a)$ to ensure only positive values: |

$$o(a) = (@resultSignificance^{o(a)})$$

2  Normalise $O$ to 100:

$$O = norm(O) \cdot 100$$

*(the results are the percentage chance of selecting a particular agent)*

3  Randomly select candidate agent $a^*$, based on its percentage chance $o(a^*)$:

Select $a^*$ where $a^* \in A \wedge random(o(a^*)) = true$

End

Following the candidate selection routine, the control goes over to the selected agent (Agent B) which is then notified with the message *MigrateOrder()*.

The agent decides about its destination node itself. It carefully examines all the nodes available for its migration. Each node is asked about its 'remaining resources' statistics [message *GetNodeStatistics()*] and is supposed to provide the appropriate information in the *NodeStatistics()* message. Note that these messages are also sent asynchronously to speed up the process:

**Algorithm**  Select an alternative node for the agent migration

Input:

All possible alternative nodes $\eta = \{n_1, n_2, ..., n_m\}$, with their 'remaining resources':

$$f(n) = \{f_1(n), f_2(n), ..., f_i(n)\}$$

$$r(t^*) = \{r_1(t^*), r_2(t^*), ..., r_i(t^*)\}$$

$@resultSignificance = 1.02$

*(this parameter is adjustable, higher values can result in never selected nodes, too low values can result in nodes selection being too random)*

Output:

Target node $n^*$, which will be the destination for the candidate agent $a^*$ migration

Begin

1  Create set $p$:

For each node $n$ in $\eta - \mu(t^*)$ :

*(current candidate agent node is not considered for migration)*

    1.1  Add $p(n)$ to $p$

    1.2  Initialise $p(n)$ to: $p(n) = \sum_{i \in \Psi}(r_i(t^*) \cdot f_i(n))$

    1.3  If $\mu(t^*) = n$ then $p(n) = p(n) \cdot c(t^*)$

    *(the candidate agents home node is favoured; the greater task migration cost the more likely agent will return to its home node)*

**Algorithm** Select an alternative node for the agent migration (continued)

| | | |
|---|---|---|
| 2 | For each point value *p(n)* in *p*: | |
| | 2.1 Process *p(n)* to ensure only positive values: | |
| | $p(n) = (@f\_\text{Re}sultSignificance^{p\ n})$ | |
| 3 | Normalise *p* to 100: | |
| | $p = norm(p) \cdot 100$ | |
| | *(the results are the percentage chance of selecting a particular node)* | |
| 4 | Randomly select a target node $n^*$, based on its percentage chance $p(n^*)$: | |
| | Select $n^*$ where $n^* \in \eta - \mu(t^*)$  $random(p\ n^*) = true$ | |

End

The candidate agent can select the best target node for migration. The agent tasks required resources are compared to the resources available on nodes. The agent can assign up to 100 points in total to all nodes but the one it is already on. If the agent was migrated before, it is going to put more pressure on its mother node and thus it assigns more points there. These values become percentage chances to select a certain node for the migration process. Please note, that agent can be migrated several times during 'system transformation'. This usually happens when there is not enough 'available resources' to serve all tasks or 'stable' configuration is impossible; in that case system will not be able solve the problem and the algorithm will return no solution after defined time.

After the successful selection of the destination node is completed, the agent migrates there. At first it detaches from the current node [*DispatchAgent()* message] and then carries its code to the destination node and recreates its state there [*AssignAgent()*].

## 4 Experiment and results

Similar to our previous system (Sliwko and Zgrzywa, 2007), a simulation has been performed with the help of JMASB. This framework has been initially developed for agent-based systems performance analysis and it enables the researcher to test even complex schemes when planning the resource management strategy. The created prototype has met the performance requirements, although the Java language is not considered a speed daemon; however, an application of various optimisations could result in its better performance (Singh and Pande, 2003).

This experiment was done with use of JMASB v1.4. Since its last presentation, the JMASB has gone under heavy development and so now it offers some extended features in comparison to its initial version. Aside from introducing new capabilities which allowed us to test the new strategy, the current JMASB version is also able to simulate some agent capabilities such as autonomy, adaptation and communication abilities, cooperation schemas, experience gathering and negotiation features (Franklin and Graesser, 1996; Goodwin, 1995; Johansson et al., 2002; Montresor et al., 2003; Xu and Wimsaldonado, 2000; Yang et al., 2005). As the JMASB has grown to a fully matured and very flexible product, we are planning to use it in our future simulations.

The system is designed to solve 'd-resource system optimisation problem' on average machine and so an accessible configuration has a considerable impact on the system performance. Thus, we decide to use low-end hardware to process, which should be widely available in our market:

**Table 1** Testing environment

| | Testing environment |
|---|---|
| Processor | Intel Pentium IV (Northwood) 2000 MHz |
| Memory | 512 MB PC2700 DDR SDRAM Kingston |
| Main board | Intel S845WD1-E (Intel 845E) |
| Java virtual machine | Java(TM) 2 runtime environment, standard edition 1.4.2_15 |
| Disc array | FastTrack100 RAID0 (Stripe 2+0) |
| Hard drive(s) | 2 × Samsung SP1614N 160GB |
| System | Windows 2000 Professional SP4 |

In Sliwko and Zgrzywa (2007), it was demonstrated that our strategy is able to handle a potentially unlimited number of resources; in this simulation we use only 2D different resources: CPU and memory. As we have already tested the system functionalities and correctness, we are going to focus on more realistic scenarios; especially we would like to compare our strategy with a few adaptations of classical and well known algorithms.

For the purpose of our experiment we have created seven different test scenarios. The goal there was not only to compare the system with other schemas, but also to test enterprise architecture features such as scalability, reliability and performance (Alur et al., 2003; Chu et al., 2006; Thai and Lam, 2002).

Every test scenario has a random agent initial assignment. Such a setup resulted in at least one overloaded node in every case and the system needed to be partially rebuilt in order to reach the 'stable system' status. Optimal solutions were computed for a few setups. However, this problem belongs to the NP-hard problems class (Cormen et al., 1990; Moura, 2002), so without the use of a supercomputer, we could only compute solutions for smaller sets.

### 4.1 Experiment configuration

The initial configuration consists of eight nodes with assigned levels of available resources. The schema has a potentially unlimited number of different kinds of resources (Sliwko and Zgrzywa, 2007). However, in this experiment we utilised only two of them, CPU and memory.

**Table 2** Experiment nodes available resources

| Node name | Available resources | |
|---|---|---|
| | CPU | Memory |
| Node 1 | 40 | 80 |
| Node 2 | 60 | 40 |
| Node 3 | 50 | 80 |

**Table 2**  Experiment nodes available resources (continued)

| Node name | Available resources | |
|---|---|---|
| | CPU | Memory |
| Node 4 | 20 | 50 |
| Node 5 | 40 | 20 |
| Node 6 | 40 | 40 |
| Node 7 | 30 | 50 |
| Node 8 | 40 | 20 |

Thirty two jobs were generated. We also defined the 'task migration cost' for each of them (please note that not every job is used in each test scenario):

**Table 3**  Experiment tasks needed resources and their 'tasks migration costs'

| Task name | Needed resources | | Task migration cost | Task name | Needed resources | | Task migration cost |
|---|---|---|---|---|---|---|---|
| | CPU | Memory | | | CPU | Memory | |
| J01 | 4 | 5 | 4 | J17 | 2 | 4 | 5 |
| J02 | 14 | 7 | 5 | J18 | 9 | 4 | 5 |
| J03 | 10 | 3 | 4 | J19 | 12 | 2 | 7 |
| J04 | 16 | 6 | 7 | J20 | 8 | 20 | 1 |
| J05 | 14 | 14 | 10 | J21 | 11 | 2 | 7 |
| J06 | 18 | 6 | 3 | J22 | 20 | 6 | 2 |
| J07 | 4 | 15 | 6 | J23 | 2 | 12 | 5 |
| J08 | 11 | 9 | 6 | J24 | 6 | 6 | 3 |
| J09 | 5 | 15 | 4 | J25 | 1 | 9 | 4 |
| J10 | 7 | 11 | 4 | J26 | 3 | 9 | 10 |
| J11 | 17 | 12 | 8 | J27 | 9 | 10 | 2 |
| J12 | 5 | 3 | 6 | J28 | 6 | 6 | 8 |
| J13 | 4 | 20 | 4 | J29 | 1 | 20 | 6 |
| J14 | 2 | 18 | 5 | J30 | 7 | 4 | 5 |
| J15 | 16 | 15 | 1 | J31 | 11 | 18 | 4 |
| J16 | 14 | 20 | 9 | J32 | 17 | 17 | 10 |

In the course of the simulation, five different strategies were deployed: FULLSCAN, IJIIDS08, KESAMSTA07, GREEDY and BALANCE.

The FULLSCAN strategy, as the name suggests, performs a full search over all available configurations. It detects working solutions and counts their 'system transformation cost'. The FULLSCAN strategy guarantees a globally optimal solution (under appropriate modelling assumptions) (Morton and Mareels, 2000). However, it is not efficient enough due to a large scale of computation level required. We performed a number of optimisations to reduce the total scope of variants to be created (Ueda, 1986), which enabled us to skip some of the unnecessary iterations. However, it still cannot be considered a real-time strategy (Reichardt, 2004; Trakhtenbrot, 1984). In our experiment, this strategy is mainly used for finding the optimal 'system transformation cost' in small setups (up to five nodes and twenty agents) and for making a comparison with the results from other strategies. In the course of the simulation, we were able to compute an optimal solution for only up to five nodes and twenty jobs. Defining more objects resulted in more than 1,018 (for six nodes and 24 agents) iterations needed to finish the task. We were not able to perform it on any computer environment available to us.

The IJIIDS08 strategy is basically what we are describing in this paper. As the strategy is not a deterministic one, we run it up to five times and compare the results.

The KESAMSTA07 strategy is an original strategy we presented in Sliwko and Zgrzywa (2007). It does not take 'tasks migration cost' into consideration, but the results are surprisingly good. The KESAMSTA07 strategy is not a deterministic one, thus it has several results presented.

The GREEDY and BALANCE strategies are the adaptations of well-known GREEDY and BALANCE approximation algorithms. Their brief descriptions can be found in Becker and Geiger (1996) and Mehta et al. (2005), respectively. Both approximation strategies were implemented as deterministic, thus we need to run them only once for each test scenario.

For each strategy a timeout counter was introduced. The strategy can work up to five minutes, with an exception of the FULLSCAN strategy which we decided to stop after 168 hours that is one full week. The FULLSCAN strategy was supposed to be a base for competence ratio, thus we decided to extend the timeout value for it. In the current JMASB implementation our experiment environment can process about 350 iterations per second on average.

## 4.2 Experiment results

The first test was meant to diagnose the correctness of the investigated strategies. In the initial setup, two nodes were used with eight jobs randomly assigned. This setup resulted in one overloaded node (*Node01*). Initially, all the overloaded nodes were *italicised* for the reader's convenience. In this case, the optimal 'system transformation' has the 'system transformation cost' which equals seven.

**Table 4.1**  The first test

| *Initial configuration* | |
|---|---|
| *Node name* | *Tasks* |
| *Node01* | J01, J02, J03, J04, J06, J07 |
| Node02 | J05, J08 |
| Node03 | (node not available during test) |
| Node04 | (node not available during test) |
| Node05 | (node not available during test) |
| Node06 | (node not available during test) |
| Node07 | (node not available during test) |
| Node08 | (node not available during test) |

Note: Initially two nodes and eight jobs were used.

**Table 4.1** The first test (continued)

| Experiment results | |
|---|---|
| *Strategy name* | *System transformation cost* |
| Highest possible migration cost (all tasks migrated) | 45 |
| FULLSCAN strategy (optimal cost) | 7 |
| IJIIDS08 strategy | 10, *7*, 11, *7*, *7* |
| KESAMSTA07 strategy | 11, *7*, 15, 8, 13 |
| GREEDY strategy | 7 |
| BALANCE strategy | 7 |

Note: Initially two nodes and eight jobs were used.

All the strategies presented the ability to compute an optimal solution in a short time. The non-deterministic algorithms generated several other solutions to the problem; among the results, were the ones we consider optimal; we marked them with the *italicised* font.

In the following sections of this paper additional results of our research are presented. Each of the table below presents one experiment with a different initial setup. We enabled one additional node and also added four agents per each setup. This resulted in one or more 'overloaded' nodes and rendered the whole system 'overloaded':

**Table 4.2** The second test

| Initial configuration | |
|---|---|
| *Node name* | *Tasks* |
| *Node01* | J01, J03, J04, J06, J10, J11 |
| Node02 | J02, J08, J09 |
| Node03 | J05, J07, J12 |
| Node04 | (node not available during test) |
| Node05 | (node not available during test) |
| Node06 | (node not available during test) |
| Node07 | (node not available during test) |
| Node08 | (node not available during test) |
| Experiment results | |
| *Strategy name* | *System transformation cost* |
| Highest possible migration cost (all tasks migrated) | 67 |
| FULLSCAN strategy (optimal cost) | *10* (strategy needed two minutes to complete task) |
| IJIIDS08 strategy | *10*, 11, 14, *10*, *10* |
| KESAMSTA07 strategy | 14, *10*, *10*, 18, 17 |
| GREEDY strategy | (GREEDY strategy could not find solution in five minutes) |
| BALANCE strategy | 11 |

In the second test, the FULLSCAN strategy determined the optimal 'system transformation cost' as ten; scanning all possible solutions, the process took less than two minutes to be completed. Then, we tested the remaining strategies. The IJIIDS08 and KESAMSTA07 algorithms needed a few seconds to find several solutions to the problem. The results included the ones that had the lowest possible 'system transformation cost' and were regarded as optimal. The GREEDY strategy did not finish after five minutes of its execution so we terminated it.

**Table 4.3** The third test

| Initial configuration | |
|---|---|
| Node name | Tasks |
| Node01 | J01, J04, J14, J16 |
| Node02 | J08, J11, J12, J15 |
| Node03 | J02, J03, J06, J07, J13 |
| *Node04* | J05, J09, J10 |
| Node05 | (node not available during test) |
| Node06 | (node not available during test) |
| Node07 | (node not available during test) |
| Node08 | (node not available during test) |
| Experiment results | |
| Strategy name | System transformation cost |
| Highest possible migration cost (all tasks migrated) | 86 |
| FULLSCAN strategy (optimal cost) | *12* (strategy needed seven hours to complete task) |
| IJIIDS08 strategy | 19, 20, 16, *12*, *12* |
| KESAMSTA07 strategy | 41, 27, 18, 44, 25 |
| GREEDY strategy | (GREEDY strategy could not find solution in five minutes) |
| BALANCE strategy | 30 |

The third test showed the difference between the IJIIDS08 and the KESAMSTA07 strategy. The KESAMSTA07 was not designed to keep the 'system transformation cost' as minimal, thus there are many differences within its results. The GREEDY strategy again could not be completed within the given timeframe.

The fourth test demonstrated the fact that we cannot use the FULLSCAN strategy for any bigger problem instances. The strategy completed its task. However, it took as many as five days, which is not acceptable in real-time systems. Both the KESAMSTA07 and IJIIDS08 strategies found good solutions. The latter one found also the optimal solution. The deterministic GREEDY strategy performed the worst, not finding any solution at all. The BALANCE strategy proposed the solution with the system transformation cost equal 47, which is not the best possible.

The fifth and sixth tests were the extension of the previous one. We could not rely on the FULLSCAN strategy for determining the optimal 'system transformation cost' value, thus we were only able to compare the results our strategies generated. All strategies finished exceptionally well. The GREEDY strategy determined a good solution, reaching the 'system transformation cost' value of 53. The BALANCE strategy resulted in the 'system transformation cost' of 77, which is an average result and exceeds the lowest computed cost by more than 70%. The best solution was computed by the IJIIDS08 strategy, with the 'system transformation cost' of 42. The KESAMSTA07 strategy was able to find the 'system transformation cost' equal 57, which is a good result in comparison to other outcomes.

**Table 4.4** The fourth test

| | |
|---|---|
| *Initial configuration* | |
| *Node name* | *Tasks* |
| *Node01* | J01, J03, J04, J05, J09, J10, J17, J18 |
| Node02 | J02, J06, J11 |
| Node03 | J08, J12, J14, J15 |
| Node04 | J07, J16 |
| *Node05* | J13, J19, J20 |
| Node06 | (node not available during test) |
| Node07 | (node not available during test) |
| Node08 | (node not available during test) |
| *Experiment results* | |
| *Strategy name* | *System transformation cost* |
| Highest possible migration cost (all tasks migrated) | 104 |
| FULLSCAN strategy (optimal cost) | *20* (strategy needed five days to complete task) |
| IJIIDS08 strategy | 21, *20*, 21, 24, 27 |
| KESAMSTA07 strategy | 30, 44, 32, 36, 47 |
| GREEDY strategy | (GREEDY strategy could not find solution in five minutes) |
| BALANCE strategy | 47 |

Note: The FULLSCAN strategy needed as much as five days to perform its task.

**Table 4.5** The fifth test

| | |
|---|---|
| *Initial configuration* | |
| *Node name* | *Tasks* |
| *Node01* | J01, J03, J04, J06, J16, J20 |
| Node02 | J08, J09, J17, J18 |
| Node03 | J07, J14, J19, J22 |
| *Node04* | J10, J11, J23, J24 |
| *Node05* | J02, J13 |
| *Node06* | J05, J12, J15, J21 |
| Node07 | (node not available during test) |
| Node08 | (node not available during test) |

Note: Please note we were not able to use FULLSCAN strategy anymore.

**Table 4.5** The fifth test (continued)

| | Experiment results |
|---|---|
| Strategy name | System transformation cost |
| Highest possible migration cost (all tasks migrated) | 121 |
| FULLSCAN strategy (optimal cost) | (strategy could not finish the task within seven days) |
| IJIIDS08 strategy | 28, 36, 36, 32, 28 |
| KESAMSTA07 strategy | 64, 50, 58, 84, 53 |
| GREEDY strategy | (GREEDY strategy could not find solution in five minutes) |
| BALANCE strategy | 57 |

Note: Please note we were not able to use FULLSCAN strategy anymore.

**Table 4.6** The sixth test

| Initial configuration | |
|---|---|
| Node name | Tasks |
| *Node01* | J03, J06, J20, J26, J28 |
| Node02 | J04, J05 |
| Node03 | J01, J17, J21 |
| *Node04* | J12, J16, J22, J24, J27 |
| *Node05* | J02, J07, J10, J13, J14, J15, J18, J23, J25 |
| Node06 | J08, J11 |
| Node07 | J09, J19 |
| Node08 | (node not available during test) |
| Experiment results | |
| Strategy name | System transformation cost |
| Highest possible migration cost (all tasks migrated) | 145 |
| FULLSCAN strategy (optimal cost) | (strategy could not finish the task within seven days) |
| IJIIDS08 strategy | 42, 55, 48, 57, 51 |
| KESAMSTA07 strategy | 78, 61, 86, 57, 64 |
| GREEDY strategy | 53 |
| BALANCE strategy | 77 |

Note: Please note the result of the GREEDY strategy.

During the final test, all nodes were enabled for a total 32 jobs setup. Here, we were not able to compute an optimal solution, however the IJIIDS08 strategy performed much better than the deterministic BALANCE algorithm, in one of its runs finding a solution with the 'system transformation cost' equal 53. The KESAMSTA07 strategy, in turn, found a working solution, however the proposed solutions were generally worse than the ones from other strategies. The GREEDY approach could not find any proper solution at all.

**Table 4.7**  The final test

| Initial configuration | |
|---|---|
| Node name | Tasks |
| Node01 | J01, J04, J16 |
| Node02 | J11, J18, J27, J28 |
| Node03 | J02, J06, J07, J19 |
| Node04 | J05, J09, J10, J17, J24 |
| Node05 | J14, J31 |
| Node06 | J08, J12, J15, J21, J25, J30 |
| Node07 | J03, J13, J20, J22, J26, J29 |
| Node08 | J23, J32 |
| Experiment results | |
| Strategy name | System transformation cost |
| Highest possible migration cost (all tasks migrated) | 170 |
| FULLSCAN strategy (optimal cost) | (strategy could not finish the task within seven days) |
| IJIIDS08 strategy | 59, 96, 53, 59, 75 |
| KESAMSTA07 strategy | 64, 128, 75, 81, 93 |
| GREEDY strategy | (GREEDY strategy could not find solution in five minutes) |
| BALANCE strategy | 86 |

Note: All nodes were enabled for total 32 jobs setup.

## 4.3 Experiment analysis

In all the tests, the tested IJIIDS08 strategy managed to find a solution to all the presented problems. The default timeout (five minutes) proved to be sufficient in all simulations. In most cases, the strategy was able to deliver a proper solution with a low 'system transformation cost'. We were able to confirm an optimal solution only in three out of seven cases, due to high computation power required.

We also tested the previously used schema – the KESAMSTA07 strategy (Sliwko and Zgrzywa, 2007). It proved to be able to return satisfying results, which, in turn, was a surprise as the KESAMSTA07 strategy was not designed for minimising the 'system transformation cost' in the first place. However, the intention of this strategy was to reduce the total algorithm iterations number, which resulted in minimising the total number of migrations. The 'task migration cost' was selected from the integer value between one and 20 as a larger scale could have probably resulted in the KESTAMSTA07 strategy being not so effective.

Surprisingly, the GREEDY strategy generated the worst results. Only in two cases was this algorithm able to find a problem solution. However, the setups used really every bit of all the available resources, thus these extreme configurations could cause this simple, but generally effective strategy (Becker and Geiger, 1996) to fail most of the tests.

The second deterministic strategy, the BALANCE algorithm, proved its usefulness in case of such problems as presented above. Generally speaking, the BALANCE is highly rated (Mehta et al., 2005) for solving problems from among the optimisation problems class (Schirmer, 1995). Another advantage of the BALANCE approach is its deterministic nature. In a real system, only one run through the problem setup is enough as it always computes the same result. Thus, it can be faster than other presented strategies which usually need more runs to find a good solution due to their random nature. On the other side, the results from the BALANCE strategy were slightly worse than from other non-deterministic strategies.

## 5 Conclusions

As far as the initial results (Sliwko and Zgrzywa, 2007) are concerned, we proved that the non-deterministic schema we used is efficient in planning a strategy for the resource utilisation in agent-based systems. Moreover, the experiment demonstrated that the presented schema tends to behave very alike in similar situations and is quite resistant to nasty setups. It was also demonstrated that such an approach can handle a potentially unlimited number of different resources without any significant performance drop.

However, the previously presented schema had certain drawbacks such as the already described 'flickering' effect, when agents were rapidly migrating to different nodes. This sometimes caused the system to waste precious iterations on purposeless migrations. Our latest research focused mainly on eliminating this disadvantage.

A few solutions were developed. The social agent behaviour strategy proved to be the best in class; its additional advantage is a significant reduction in the number of iterations needed in the searching for solution process. A simple pattern of 'task migration cost' was introduced. It serves now as an objective function (Moura, 2002) which quantifies the optimality (Cormen et al., 1990; Goldberg, 1989) of a proposed solution.

Agent ability to cooperate and adapt (Goodwin, 1995) also proved to be very useful. Agents are now cost conscious and they have the autonomy to decide about their actions. Basically, an agent does not 'want' to be migrated and a migrated agent 'tries' to return to its base node. This approach eliminated the 'flickering' effect almost completely.

The presented strategies were also tested against the optimal solution which was computed with the help of the full scanning (Trakhtenbrot, 1984) pattern, sometimes named also the brute force (Morton and Mareels, 2000) or the exhaustive search (Ueda, 1986) approach. This experiment demonstrated that the algorithm meets its requirements and the presented strategy proved competitive enough.

It is necessary to mention another drawback of the KESAMSTA07 strategy (Sliwko and Zgrzywa, 2007), when nodes were not able to serve all requests. In our current work we did not focus on detecting and handling such situations. Due to being decentralised, our system needs a number of messages to update the load information and the balance load among the nodes. In this respect, our system needs some further improvement. In order to address both issues, we are considering the introduction of a node-hierarchy structure and more sophisticated negotiation schemas in our future work on the system.

# References

Alur, D., Malks, D. and Crupi, J. (2003) *Core J2EE Patterns: Best Practices and Design Strategies*, 2nd ed., Prentice Hall.

Becker, A. and Geiger, D. (1996) 'Optimization of Pearl's method of conditioning and greedy-like approximation algorithms for the vertex feedback set problem', *Artificial Intelligence*, Vol. 83, No. 1, pp.167–188.

Buyya, R. (1999) *High Performance Cluster Computing: Architectures and Systems*, Vols. 1 and 2, Prentice Hall.

Cabri, G., Ferrari, L., Leonardi, L. and Quiadamo, R. (2006) *Strong Agent Mobility for Aglets Based on the IBM JikesRVM*, Universita di Modena e Reggio Emilia, Dipartimento di Ingegneria dell'Informazione, pp.91–93.

Chu, K., Cordero, O., Korf, M., Pickersgill, C., Whitmore, R., Bathurst, W., Dobrik, R., Kodali, R. and Zimmermann, R. (2006) *Oracle® SOA Suite Developer's Guide 10g (10.1.3.1.0)*, B28764-01, Oracle.

Cormen, T-H., Leiserson, C-E., Rivest, R-L. and Stein, C. (1990) 'Introduction to algorithms', *MIT Electrical Engineering and Computer Science*.

Craswell, N-E., Haines, J., Humphreys, B., Johnson, C. and Thistlewaite, P. (1997) 'Aglets: a good idea for spidering?', *4th IDEA Workshop Proceedings*.

Denning, J-P. (1976) 'Fault tolerant operating systems', *ACM Computing Surveys*, Vol. 8, No. 4, pp.359–389.

Dillenseger, B. and Hazard, L. (2003) 'A dynamic distribution and load balancing experiment with synchronous programming-based mobile objects', *Lecture Notes in Computer Science*, Vol. 2888, pp.1188–1207.

Epstein, L. and van Stee, R. (2005) 'Optimal online algorithms for multidimensional packing problems', *SIAM Journal on Computing Archive*, Vol. 35, No. 2, pp.431–448.

Franklin, S. and Graesser, A. (1996) 'Is it an agent or just a program?: a taxonomy for autonomous agents', *Lecture Notes in Computer Science*, Vol. 1193, pp.21–35.

Garey, M-R. and Johnson, D-S. (1979) *Computers and Intractability: A Guide to the Theory of NP-Completeness*, W.H. Freeman & Co.

Goldberg, D-E. (1989) *Genetic Algorithms in Search, Optimization and Machine Learning*, Addison-Wesley Professional.

Goodwin, R. (1995) 'Formalizing properties of agents', *Journal of Logic and Computation*, Vol. 5, No. 6, pp.763–781.

He, J. (2000) 'An architecture for wide area network load balancing', *Proceedings of the 2000 IEEE International Conference on Communications*, pp.1169–1173.

Heineman, G-T. and Councill, W-T. (2001) *Component-Based Software Engineering: Putting the Pieces Together*, Addison-Wesley Professional.

Ho, K-S. and Leong, H-V. (2000) 'A multi-agent negotiation algorithm for load balancing in CORBA-based environment', *Lecture Notes in Computer Science*, Vol. 1983, pp.314–319.

Ibaraki, T. and Yagiura, M. (2004) 'Recent metaheuristic algorithms for the generalized assignment problem', *Proceedings of the International Conference on Informatics Research for Development of Knowledge Society Infrastructure*, pp.229–237.

Johansson, S., Davidsson, P. and Kristell, M. (2002) 'Four multi-agent architectures for intelligent network load management', *Lecture Notes in Computer Science*, Vol. 2521, pp.239–248.

Johansson, S., Davidsson, P. and Kristell, M. (2003) 'Cooperative negotiation in a multi-agent system for real-time load balancing of a mobile cellular network', *Proceedings of the Second International Joint Conference on Autonomous Agents and Multiagent Systems*, pp.568–575.

Jones, J. and Crickell, C. (1997) *Second Evaluation of Job Queuing/Scheduling Software*, Tech. report NAS-97-013, NASA Ames Research Center, pp.1–34.

Kesselman, F. and Kesselman, C. (1999) *The Grid: Blueprint for a New Computing Infrastructure*, Morgan Kaufmann Publishers.

Kim, G-S., Kim, K. and Eom, Y-I. (2004) 'Dynamic load balancing scheme based on resource reservation for migration of agent in the pure P2PNetwork environment', *Lecture Notes in Computer Science*, Vol. 3397, pp.538–546.

Mehta, A., Saberi, A., Vazirani, U., Vazirani, V. and Mehta, A. (2005) 'AdWords and generalized on-line matching', *Proceedings of 46th Annual IEEE Symposium on Foundations of Computer Science*, pp.264–273.

Montresor, A., Meling, H. and Babaoglu, O. (2003) 'Messor: load-balancing through a swarm of autonomous agents', *Lecture Notes in Artificial Intelligence*, Vol. 2530, pp.125–137.

Morton, A-B. and Mareels, I.M.Y. (2000) 'An efficient brute-force solution to the network reconfiguration problem', *IEEE Transactions on Power Delivery*, Vol. 15, No. 3, pp.996–1000.

Moura, L. (2002) 'Introduction to the theory of NP-completeness', *University of Ottawa Lectures*, Vols. 1–5, pp.1–66.

Nguyen, N-T. and Sliwko, L. (2005) 'Zastosowanie systemu wieloagenckiego do wyszukiwania informacji w sieci internet', *Proceedings of IWSE 2006*, pp.155–164.

Nguyen, N-T., Gandza, M. and Paprzycki, A. (2006) 'A consensus-based multi-agent approach for information retrieval in internet', *Proceedings of ICCS 2006, Lecture Notes in Computer Science*, Vol. 3993, pp.208–215.

Parnas, D. (1972) 'On the criteria to be used in decomposing systems into modules', *Communications of the ACM*, pp.1053–1058.

Qu, X., Yang, X., Gui, C. and Fan, W. (2004) 'A policy-based service-oriented grid architecture', *Lecture Notes in Computer Science*, Vol. 3033, pp.597–603.

Randell, B., Lee, P. and Treleaven, P-C. (1978) 'Reliability issues in computing system design', *ACM Computing Surveys*, Vol. 10, No. 2, pp.123–165.

Reichardt, B-W. (2004) 'The quantum adiabatic optimization algorithm and local minima', *Proceedings of the 36th Annual ACM Symposium on Theory of Computing*, pp.502–510.

Schirmer, A. (1995) *A Guide to Complexity Theory in Operations Research*, Manuskripte aus den Instituten fur Betriebswirtschaftslehre der Universitat Kiel.

Shi, D., Yin, J., Zhang, W., Dong, J. and Xiong, D. (2006) 'A distributed collaborative design framework for multidisciplinary design optimization', *Lecture Notes in Computer Science*, Vol. 3865, pp.294–303.

Singh, A. and Pande, S. (2003) 'Compiler optimizations for Java Aglets in distributed data intensive applications', *Proceedings of the 2002 ACM Symposium on Applied Computing*, pp.87–92.

Sliwko, L. and Nguyen, N-T. (2007) 'Using multi-agent systems and consensus methods for information retrieval in internet', *International Journal of Intelligent Information and Database Systems*, Vol. 1, No. 2, pp.181–198.

Sliwko, L. and Zgrzywa, A. (2007) 'Multi-resource load optimization strategy in agent-based systems', *Lecture Notes in Artificial Intelligence*, Vol. 4496, pp.348–357.

Szyperski, C., Gruntz, D. and Murer, S. (2002) *Component Software: Beyond Object-Oriented Programming*, 2nd ed., Addison-Wesley Professional.

Thai, T-L. and Lam, H. (2002) *.Net Framework Essentials*, O'Reilly Programming Series.

Trakhtenbrot, B-A. (1984) 'A survey of Russian approaches to Perebor (brute-force searches) algorithms', *IEEE Annals of the History of Computing*, Vol. 6, No. 4, pp.384–400.

Ueda, K. (1986) 'Making exhaustive search programs deterministic', *Proceedings of 3rd International Conference on Logic Programming*, pp.270–282.

Wang, X., Hallstrom, J. and Baumgartner, G. (2001) *Reliability Through Strong Mobility*, Technical report, Department of Computer and Information Science, The Ohio State University, pp.2–5.

Woeginger, G-J. (2003) 'Exact algorithms for NP-hard problems: a survey', *Lecture Notes in Computer Science*, Vol. 2570, pp.185–207.

Wurz, M. and Schuldt, H. (2005) 'Dynamic parallelization of grid-enabled web services', *Lecture Notes in Computer Science*, Vol. 3470, pp.173–183.

Xu, C-Z. and Wimsaldonado, B. (2000) 'A mobile agent based push methodology for global parallel computing', *Java Grande 1999, Special Issue of Concurrency: Practice and Experience*, pp.2–12.

Yagiura, M. and Ibaraki, T. (2001) 'On metaheuristic algorithms for combinatorial optimization problems', *Systems and Computers in Japan*, Vol. 32, No. 3, pp.33–55.

Yang, Y., Chen, Y., Cao, X. and Juguyen, J. (2005) 'Load balancing using mobile agent and a novel algorithm for updating load information partially', *Lecture Notes in Computer Science*, Vol. 3619, pp.1243–1252.

Yoshiharu, K., Keidi, F-M., Prabhakar, R. and Yoshiko, W. (2004) 'Multidimensional cube packing', *Algorithmica*, Vol. 40, No. 3, pp.173–187.